\documentclass[aps,prl,twocolumn,showpacs,10pt]{revtex4-1}%
\usepackage{graphicx}
\usepackage[usenames]{color}
\usepackage{amsmath}
\usepackage{amsfonts}
\usepackage{amssymb}%
\newcommand{\buck}{$\text{C}_{60}$}
\newcommand{\rug}{$\text{C}_{70}$}

\newcommand{\vect}[1]{{\mathbf #1}}

\begin{document}
\title{Fusion mechanism in fullerene-fullerene collisions\\
  --The deciding role of giant oblate-prolate motion
}
\author{J. Handt}
\author{R. Schmidt}
\thanks{Corresponding author}
\email{Ruediger.Schmidt@tu-dresden.de}
\affiliation{Institut f\"{u}r Theoretische Physik, Technische Universit\"{a}t Dresden,
D-01062 Dresden, Germany}
\date{\today }

\begin{abstract}
We provide answers to long-lasting questions in the puzzling behavior of fullerene-fullerene fusion:
Why are the fusion barriers so exceptionally high and the fusion cross sections so extremely small?
An \emph{ab initio} nonadiabatic quantum molecular dynamics (NA-QMD) analysis of \buck{}+\buck{} collisions reveals that the dominant excitation of an exceptionally ``giant'' oblate-prolate $H_g(1)$ mode plays the key role in answering both questions.
From these \emph{microscopic} calculations, a \emph{macroscopic} collision model is derived, which
reproduces the NA-QMD results.
Moreover, it predicts \emph{analytically} fusion barriers for different fullerene-fullerene combinations in excellent agreement with experiments.
\end{abstract}

\maketitle

\section{Introduction}
After the pioneering \buck{}+\buck{} collision experiment of E.E.B. Campbell et.al. in 1993~\cite{Campbell93},
 cluster-cluster collisions became a versatile new field of research 
 (for reviews see~\cite{Schmidt1995,Jellinek99,Campbell00_coll,Campbell2003}) with lasting interest~\cite{Schmidt91,Seifert91,Knospe1993,Schmidt1993,Schmidt92,Schmidt94,Schulte95,Zhang1994,Shen95,Farizon1999,Braeuning03,Rogan2004,Kamalou2006,Alamanova2007,Jakowski2012}.
In particular, fusion between fullerenes has been studied in great detail, both experimentally and theoretically~\cite{Campbell93,Yeretzian92,Strout93,Zhang93,Kim1994,Robertson95,Xia96,Rohmund96,Knospe96,Rohmund96_jpb,Xia97,Glotov01,Campbell2002,Zhao2002,Han2004,Kaur2008,Jakowski2010,Zettergren2014}.
Fusion is a universal phenomenon in collisions between complex particles
covering many orders in size and energy from heavy ion collisions in nuclear physics~\cite{Bock1980} to macroscopic liquid droplets~\cite{Brenn1989,Menchahca1997} or even colliding galaxies~\cite{Struck1999}.
It is a great challenge of ongoing interest to reveal universal similarities and basic differences of these mechanisms.

Usually, the gross features of fusion can be understood with macroscopic arguments~\cite{Schmidt92,Bock1980,Menchahca1997,Hasse1988} leading to the general expression for the fusion cross section $\sigma$ as function of the center mass energy $E_\text{c.m.} \equiv E$ of 
\begin{equation}
  \sigma(E)= \pi R_{\scriptscriptstyle 12}^2 \left( 1- \frac{V_B}{E} \right) \label{eq:sigma}
\end{equation}
with $R_{\scriptscriptstyle 12} = R_{\scriptscriptstyle 1}+R_{\scriptscriptstyle 2}$ the sum of the radii of the colliding partners and $V_B$ the fusion barrier (for a derivation see e.g.~\cite{Jellinek99,Bock1980,Levine1987}).
This formula (known as ``critical distance model'' in nuclear physics~\cite{Bock1980} or ``absorbing sphere model'' in chemistry~\cite{Levine1987}) describes \emph{quantitatively} the experimental fusion cross section for atomic nuclei (with $V_B > 0$) and liquid droplets (with $V_B = 0$)~\cite{Schmidt92,Brenn1989,Menchahca1997} in the low energy range with $E \gtrsim V_B$~\footnote{We do not consider here the high energy range, where the cross sections generally decreases with $E_\text{c.m.}$, see~\cite{Schmidt92}.}.
It is expected to hold also for collisions between metallic clusters~\cite{Schmidt92} (even with
$V_B< 0$); see last paper in the series~\cite{Schmidt91,Seifert91,Knospe1993,Schmidt1993}.
The physics behind formula~(\ref{eq:sigma}) is indeed simple: fusion takes place, if the colliding partners  touch, owing to the larger binding energy of the fused compound~\cite{Hasse1988}.
For colliding fullerenes one naturally expects a fusion barrier $V_B$ of at least the $sp^2$ bond
breaking energy, or more pertinently the Stone-Wales transformation energy~\cite{Zhao2002,Han2004} of a few~eV~\cite{Dresselhaus} and, according to~(\ref{eq:sigma}),
a fusion cross section of the order of the geometrical one, $\pi R_{\scriptscriptstyle 12}^2 \sim 150$~\AA{}$^2$~\cite{Rohmund96_jpb}.
Experimentally, however, the fusion barriers are about one order of magnitude larger (around 80~eV)~\cite{Rohmund96_jpb} and the cross sections are even two magnitudes smaller (a few \AA{}$^2$)~\cite{Glotov01}.

Up to now, there is no definite explanation for these findings, albeit some possible phenomenological
reasons have been discussed~\cite{Jellinek99,Campbell00_coll,Campbell2003,Glotov01,Campbell2002}.
In addition, previous (at that time still approximate) microscopic Quantum Molecular Dynamics (QMD) calculations predicted the large fusion barriers~\cite{Knospe96}.
From these studies it is also well known that only very few mutual initial orientations of the colliding cages lead to fusion (without identifying them).
Anyway, fullerenes typically do {\bf not} fuse if they touch, even at high impact energies, and the question remains, why?
In this work, we provide a clear answer to this longstanding question.

\section{Microscopic results}
Motivated by our recent findings of the dominating role of the $A_g(1)$ breathing mode in \buck{}-laser interaction~\cite{Laarmann07,Fischer2013}, we have reanalyzed fullerene-fullerene collisions with the help of the \emph{ab initio} nonadiabatic quantum molecular (NA-QMD) method~\cite{1Kunert03,Fischer2014_1,Fischer2014_2,Fischer2014_3}.
For systems as large as we are investigating here, NA-QMD is numerically more efficient than 
its \emph{ab initio} QMD approximation~\cite{Fischer2014_1,Fischer2014_2,Fischer2014_3}.
In extension to previous studies~\cite{Jellinek99,Rohmund96,Knospe96,Rohmund96_jpb,Glotov01} we include a normal mode analysis~\cite{Zhang04} of the vibrational kinetic energy.
This method decomposes the total kinetic vibrational excitation energy $E_\text{vib}$ into the individual contributions of all 174 eigenmodes of C$_{60}$ as function of time $t$ according
to
\begin{equation*}
 E_\text{vib}(t) = \sum_{i=1}^{60} \frac{m_c}{2} \dot{\vect{r}}_i^2  = \sum_{\nu=1}^{174}\frac{m_c}{2} \left( \sum_{i=1}^{60} \dot{\vect{r}}_i \vect{b}_{i\nu} \right )^2
\end{equation*}
with the atomic carbon mass $m_c$, the atomic velocities $\dot{\vect{r}}_i$ (in the molecular center of mass system without rotational components) and the normal mode eigenvectors $\vect{b}_{i\nu}$.

\begin{figure}
  \includegraphics[width=0.49\columnwidth]{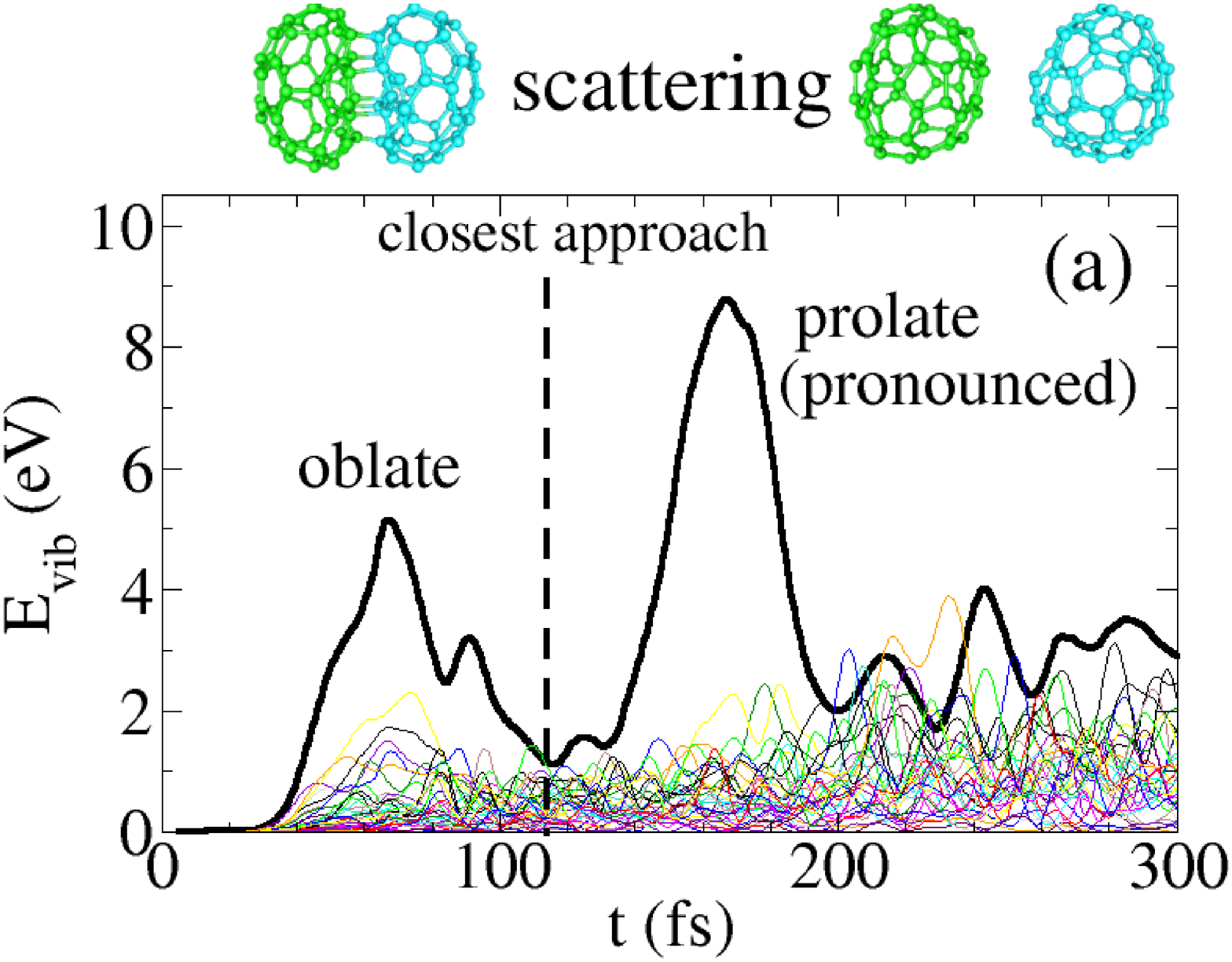}
  \includegraphics[width=0.49\columnwidth]{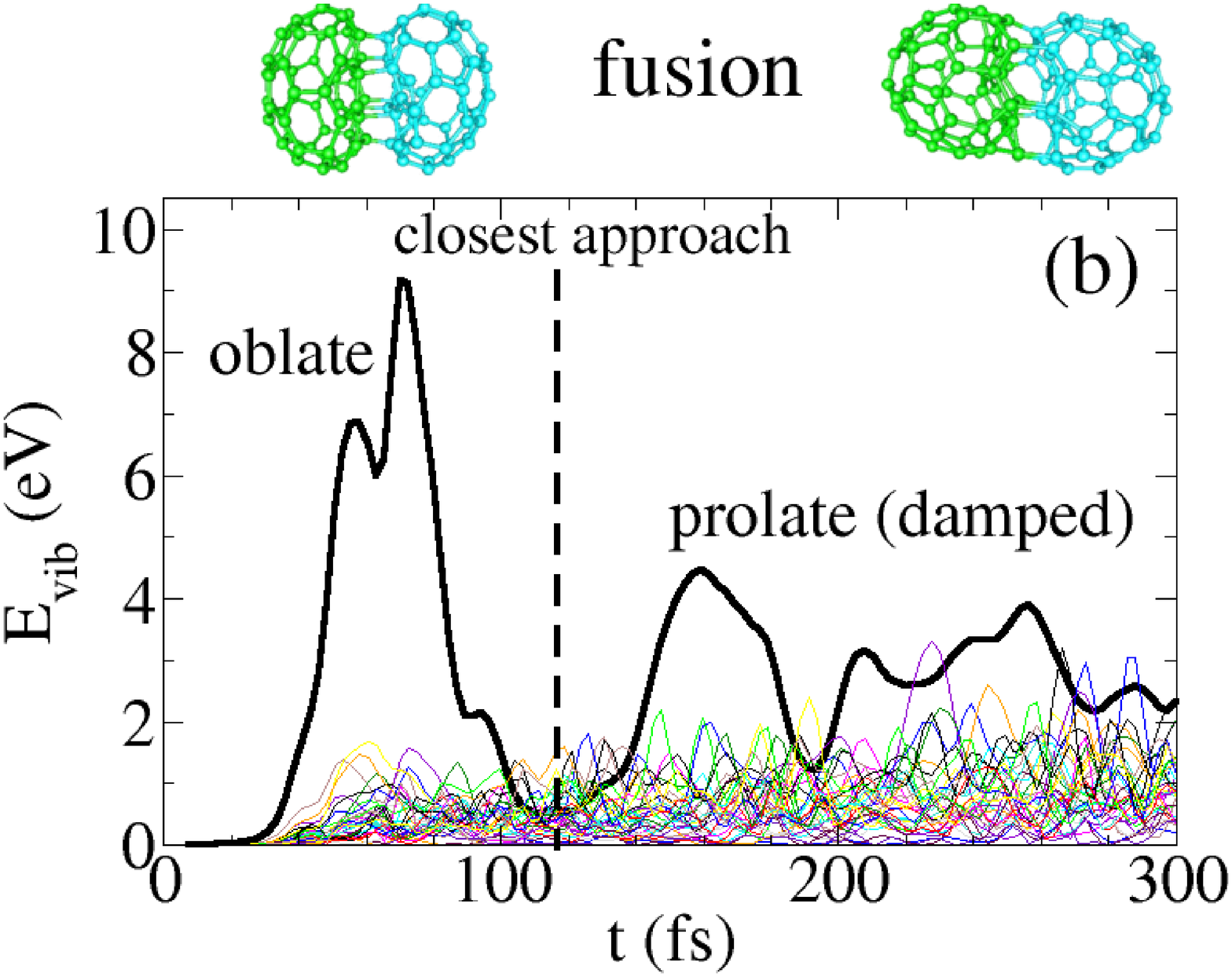}
  \caption{(color online) Normal mode analysis of central \buck{}+\buck{} collisions at an impact energy of $E_\text{c.m.}=104$~eV.
    Shown are the vibrational kinetic energies $E_\text{vib}$ of all 174 internal normal modes as function of time $t$ for (a) typical scattering and (b) typical fusion events.
  Distances of closest approach $R^\text{ret}$ are indicated by vertical, dashed lines.
  The dominating role of the $H_g(1)$ mode (black thick lines) is apparent.    
	  \label{fig:vib_analysis}}
\end{figure}

In fig.~\ref{fig:vib_analysis}, such an analysis, is shown for two central \buck{}+\buck{} collisions with the same impact energy of $E_\text{c.m.} = 104$~eV but for different initial orientations
of the clusters, leading in one case to scattering (fig.~\ref{fig:vib_analysis}(a)) and, in the other, to fusion (fig.~\ref{fig:vib_analysis}(b)).
In both cases, the extraordinary dominance of the $H_g(1)$ mode is obvious.

\begin{figure}
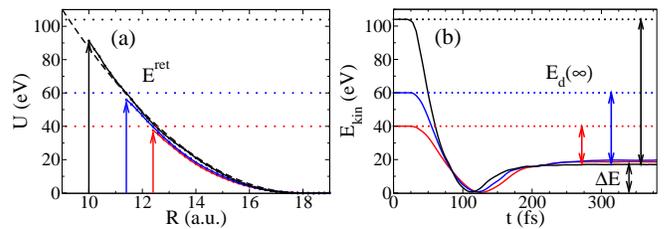

    \begin{minipage}{\columnwidth}
	    \includegraphics[width=0.49\columnwidth]{fig2a.eps}
	    \includegraphics[width=0.49\columnwidth]{fig2b.eps}
    \end{minipage}
    \caption{(color online) Typical NA-QMD results for central \buck+\buck~collisions at impact energies
     of  $E_\text{c.m.} = 40$~eV (red), $60$~eV (blue), and $104$~eV (black), indicated as dotted
     horizontal lines, and obtained by ensemble averaging over 20 different initial orientations.
    (a): The potential (deformation) energy of the relative motion $U(R)$ in the entrance channel as function of the distance $R$.
    The locations and lengths of the vertical arrows denote the distances of closest approach $R^\text{ret}$ and the total deformation energy $E^\text{ret} \equiv U(R^\text{ret})$ stored at $R^\text{ret}$, respectively.
    The dashed line is the harmonic fit $U(R)= \frac{k}{2} \bigl( R-R^{\scriptscriptstyle 0} \bigr)^{\scriptscriptstyle 2}$ with $k = 0.12$~a.u. and $R^0= 18.5$~a.u.
    (b): The kinetic energy of the relative motion $E_{kin}$ as function of time $t$ for scattering events.
    The mean values of the final kinetic energies $\Delta E \approx 17$~eV and of the corresponding
    energies dissipated into internal DOF $E_d(\infty)=E_\text{c.m.}-\Delta E$ are indicated by double arrows.
    \label{fig:qmd}}
\end{figure}

During approach, this mode is very strongly excited to a much higher degree than any other vibrational mode.
Its excitation energy of several eV is huge as compared to a single quantum of this mode of $\hbar \omega_\text{mode} \sim 34$~meV~\cite{Dong93} and, thus, its amplitude
is extremely large, as compared to a typical displacement of the elementary excitation (``giant'' $H_g(1)$ mode).
Consequently 
at the distance of closest approach $R^\text{ret}$, a highly deformed \emph{oblate-oblate} configuration of the double cluster system is formed (see above illustrations in fig.~\ref{fig:vib_analysis}).
This clearly distinguishable state accommodates practically the whole impact energy into deformation (potential) energy, which is quantitatively shown in fig.~\ref{fig:qmd}(a), also for other impact energies.
Up to this stage of the collision, there is no appreciable difference to the other collision systems (nuclei, droplets), where at this ``critical distance'' the system loses immediately its memory and
the energy is dissipated into internal degrees of freedom (DOF) leading to a hot compound. 

The fundamental difference to the other systems consists in the specific properties of the oblate-prolate mode in fullerenes and its special role in collisions.
First of all, among all vibrational modes, the $H_g(1)$ mode in \buck{} has the largest oscillation period of $T = 122$~fs~\cite{Dong93}.
This is comparable with a typical collision time and therefore, once excited, this mode will \emph{not} lose immediately its memory, as clearly seen in figs.~\ref{fig:vib_analysis}(a) and~\ref{fig:vib_analysis}(b).
Second, the oblate-prolate mode is the only eigenmode which can couple directly to the relative motion via its elongated \emph{prolate-prolate} configuration, provided it survives the dissipative coupling to all the other internal modes.
This is exactly what happens in \buck{}+\buck{} collisions and results in the majority of cases in scattering, like a \emph{fission} process with the \emph{prolate-prolate} configuration at the scission point~\cite{Hasse1988} (note the pronounced excitation of the mode in fig.~\ref{fig:vib_analysis}(a) at $t \sim 165$~fs).
Only strong coupling to the internal DOF can prevent this mechanism, allowing for fusion (note the strongly damped prolate oscillation of the mode in fig.~\ref{fig:vib_analysis}(b) at $t \sim 165$~fs).
Thus, the competitive coupling of the oblate-prolate mode to the relative motion and to all the other (bath-like) vibrational DOF determines the reaction channel ! 

The strength of the bath coupling is solely determined by the amount of energy stored in the mode at $R^\text{ret}$.
This coupling  dominates, if the energy exceeds a certain limit which generally can happen only at appropriate large impact energies.
This explains (at least preliminarily and qualitatively) the high fusion barriers.
At a fixed impact energy just above the barrier $E_\text{c.m.} \gtrsim V_B$ (as in fig.~\ref{fig:vib_analysis}), only very rare and specific initial orientations of the clusters can lead to high $H_g(1)$ excitation energies , namely, those with the principal axes of the $H_g(1)$ mode aligned to the collision axis, ensuring maximal energy transfer, which is the case in the example shown in fig.~\ref{fig:vib_analysis}(b)
(note that $E_\text{vib}$ of the mode during approach in the case of fusion (fig.~\ref{fig:vib_analysis}(b)) is twice as large as compared to scattering (fig.~\ref{fig:vib_analysis}(a))).
This, finally, explains the low fusion cross sections and completes the present, new picture of the fusion mechanism.

It modifies also the hitherto existing interpretation of scattering, as a ``bouncing off'' mechanism~\cite{Jellinek99,Campbell00_coll,Strout93,Knospe96,Glotov01}, like in collisions between two soccer balls.
Instead, ``fission'' via the prolate-prolate configuration strongly suggests, that the final kinetic energy of the fragments is largely independent on impact energy.
This is nicely confirmed in the calculations and shown in fig.~\ref{fig:qmd}(b).

The collision scenario presented in figs.~\ref{fig:vib_analysis} and~\ref{fig:qmd} is characteristic for fullerene-fullerene reactions and qualitatively observed in our NA-QMD calculations also for the other combinations and finite, small impact parameters.
Despite its microscopic complexity, the mechanism is nevertheless simple and can be understood and described by ordinary \emph{macroscopic} concepts, as will be shown in the following.
\begin{figure}
    \begin{minipage}{0.9\columnwidth}
	\begin{minipage}{\columnwidth}
	    \includegraphics[width=\columnwidth]{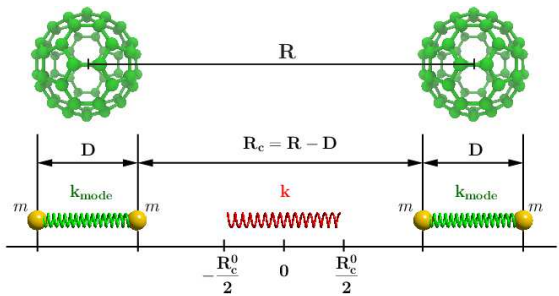}
       \end{minipage}	    
    \end{minipage}
    \caption{(color online) Phenomenological collision model with two DOF, the distance between the centers $R(t)$ and the diameter $D(t)$ of the fullerenes:
    Two collinear colliding springs (with spring constants $k_\text{mode}$ and masses $m$) interact during approach via a third, massless spring (spring constant~$k$ and initial length~$R_c^0$) located at the collision center. \label{fig:springs}}
\end{figure}

\section{Collision model}
\begin{figure}
    \begin{minipage}{\columnwidth}
	\begin{minipage}{\columnwidth}
	    \includegraphics[width=\columnwidth]{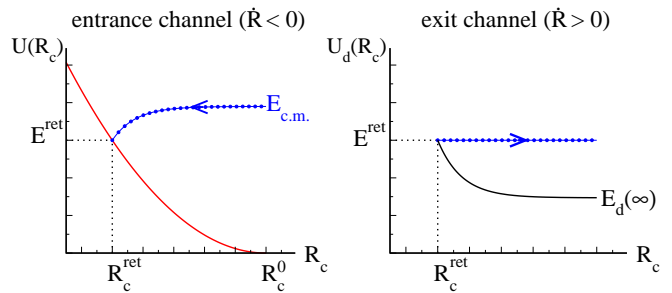}
        \end{minipage}	    
    \end{minipage}
    \caption{(color online) Schematic plot of the potential energy in the entrance channel $U(R_c)$ (left) and, for the case of scattering, in the exit channel $U_d(R_c)$ (right) as function of the contact distance $R_c$.
      Dotted blue lines with directional arrows indicate a typical trajectory (idealized as a straight line in the exit channel) with impact energy $E_{c.m.}$ at the initial distance $R_c^0$ and potential energy $E^\text{ret}$ at the distance of closest approach $R_c^\text{ret}$.
      The asymptotic value of the potential energy in the exit channel $U_d(R_c \rightarrow \infty) \equiv E_d(\infty)$ is also indicated.
    \label{fig:model}}
\end{figure}

The basic idea is to reduce drastically the {360-dimensional} scattering problem to a one-dimensional
one with only two, but relevant collective DOF, treated explicitly in a time-dependent fashion: the distance between the centers $R(t)$ (relative motion) and the diameters of the fullerenes $D(t)$ (aligned along their principal axes of the oblate-prolate mode).
Both are coupled via the contact distance $R_c = R - D$~(see fig.~\ref{fig:springs}).
The coupling to the other internal DOF will be treated implicitly in the exit channel only.
The macroscopic model is designed as follows: 

(i) In the entrance channel, the system consists of two collinear colliding springs with initial lengths $D(t=0)=D^0$, spring constants $k_\text{mode}$ and masses $m = \frac{M}{2}$ at the
ends (describing the fullerenes with mass $M$ and their $H_g(1)$ modes).
Tightly located in between there is a third, massless spring with initial length $R_c^0$ and constant $k$, describing the repulsive potential $U$ during approach (remember fig.~\ref{fig:qmd}(a) and see fig.~\ref{fig:springs} and fig.~\ref{fig:model}(a)). 

(ii) In the exit channel, the massless repulsive spring is replaced by a ``dissipative'' potential $U_d$,
which describes the coupling to all other internal vibrational DOF, and hence, controls the reaction channel (see fig.~\ref{fig:model}(b)).

The coupled Newton equations in the entrance channel read
\begin{equation} \label{equ:newton}
  \begin{split}
         \mu \ddot R &= \left . -\frac{d U}{d R_c}\right |_{R_c=R-D}\\
      \mu_\text{mode} \ddot D &= -\frac{d V_\text{mode}}{d D} + \frac{1}{2}\ \left . \frac{d U}{d R_c}\right |_{R_c=R-D}
  \end{split}
\end{equation}
with the reduced masses $\mu =\frac{M}{2}$ and the mass constant $\mu_\text{mode}=\frac{M}{4}$.
The harmonic potentials are given by 
$U(R_c) = \frac{k}{2} (R_c-R_{c}^0)^2$ with $k=0.12$~a.u. (see fig.~\ref{fig:qmd}(a))
and $V_\text{mode}(D) = \frac{k_\text{mode}}{2} (D-D^0)^2$ with the spring constant $ k_\text{mode} = \mu_\text{mode}\ \omega_\text{mode}^2=0.51$~a.u., obtained from the experimental frequency of the $H_g(1)$ mode in \buck{} ($\omega_\text{mode}=273\ \text{cm}^{-1}$~\cite{Dong93}).
The equations of motion (EOM)~(\ref{equ:newton}) can be solved analytically (see appendix).
They describe the collision up to the distance of closest approach,
i.e., the classical returning point $R_c^\text{ret}$. 
At this point the potential $U(R_c)$ for $R_c >  R_c^\text{ret}$ is replaced by the ``dissipative'' one $U_d(R_c)$ which, in dependence on $E_\text{c.m.}$, controls the outcome.
It is therefore repulsive (leading in any case to scattering) or attractive
(leading usually to fusion, but not necessarily always).
Thus, it has the general form
\begin{align}
  U_d(R_c) &= \left( E^\text{ret} - E_d(\infty) \right) f(R_c-R_c^\text{ret}) + E_d(\infty) \label{equ:pot_diss}
\end{align}
where the form factor $f$ must fulfill the conditions, $f(0)=1$ (ensuring the continuity of the potential at $R_c^\text{ret}$) and $f(\infty)=0$ (making sure that the maximal amount of dissipated energy  cannot exceed $E_d(\infty)$, in the case of scattering).
In fact, the concrete radial dependence of the potential~(\ref{equ:pot_diss}) is not relevant, and thus, we choose a simple exponential form  of $f(x)=\exp{\bigl(-\frac{x}{\Delta}\bigr)}$ 
with $\Delta =\bigl | \frac{R_c^\text{ret}-R_{c}^0}{2}\ (1-\frac{E_d(\infty)}{E^\text{ret}}) \bigr|$,  which guarantees also continuity of the force at $R_c^\text{ret}$ in the case of scattering.

\begin{figure}[t]
     \includegraphics[width=\columnwidth]{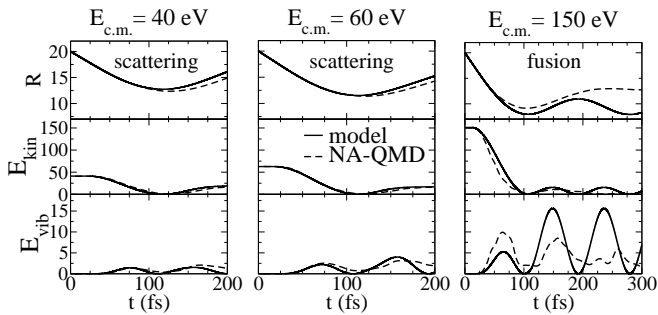}
    \caption{Comparison between NA-QMD (dashed lines) and model calculations (solid lines)
      for central \buck+\buck~collisions at three impact energies $E_\text{c.m.}$:
      distance between the centers $R$ in a.u. (top),
      kinetic energy of the relative motion $E_{kin}$ (middle),
      and vibrational kinetic energy of the $H_g(1)$ mode $E_{vib}$ (bottom) in eV
      as function of time $t$.
    \label{fig:compare_springs_exp}}   
\end{figure}

With this, all model parameters ($k$, $k_\text{mode}$, $\Delta E$) are fixed and the EOM~(\ref{equ:newton}) can be easily solved numerically.
The results are shown in fig.~\ref{fig:compare_springs_exp} and compared with NA-QMD calculations (for movies see \emph{www.dymol.org}).

The model reproduces nearly precisely the microscopic calculations for both scattering ($E_\text{c.m.} = 40, 60$~eV) and fusion ($E_\text{c.m.} = 150$~eV).
The ongoing oscillations of some quantities in the exit channel are the natural consequence of the absence of a damping mechanism for the $H_g(1)$ mode in the spring model.
The most impressive result, however, concerns the fusion barrier predicted by the model of $V_B = 85$~eV,
which is in excellent agreement with former (extremely expensive) QMD calculations~\cite{Knospe96} of $V_B = 80$~eV.

\section{Analytical solution for the fusion barrier}
The simplicity of the EOM~(\ref{equ:newton}) and the transparency of the ansatz~(\ref{equ:pot_diss})
encouraged us to further elaborate the model and to demonstrate its predictive power.
As mentioned already, the EOM~(\ref{equ:newton}) can be solved analytically, and, from this solution one can derive an approximate expression for the energy stored during approach $E^\text{ret}$ (see Fig~\ref{fig:model}(a)),
resulting in a linear dependence on the impact energy of $E^\text{ret} = \alpha( \kappa)\ E_\text{c.m.}$ with $\kappa$ the ratio of the spring constants
\begin{align}
  \kappa &= \frac{k_\text{mode}}{k} =  \frac{\frac{M}{4} \omega_\text{mode}^2}{k} \label{equ:kappa}
\end{align}
and the universal function $\alpha(\kappa)$ of
\begin{align}
      \alpha(\kappa) &= \frac{1}{4 (\kappa^2+1)}\ \left[ \frac{\kappa-1-\sqrt{\kappa^2+1}}{\sqrt{\kappa+1+\sqrt{\kappa^2+1}}} \right. \notag
		    \\ & \times \left. \sin{(\frac{\pi}{2}\frac{\kappa+1+\sqrt{\kappa^2+1}}{\sqrt{2\kappa}})}
		     - \frac{\kappa-1+\sqrt{\kappa^2+1}}{\sqrt{\kappa+1-\sqrt{\kappa^2+1}}} \right]^2		     
\label{equ:alpha}
\end{align}
(see appendix).
The difference $(E^\text{ret} - E_d(\infty))$ determines the (positive or negative) slope of the potential $U_d$~(\ref{equ:pot_diss}) at the returning point $R_c^\text{ret}$.
Taking an idealized trajectory as shown in fig.~\ref{fig:model} (i.e., neglecting the acceleration of the mode near the barrier), the fusion condition simply reads $E^\text{ret} \stackrel{!}{=} E_d(\infty)$ at $E_\text{c.m.} = V_B$.
With $E_d(\infty)=E_\text{c.m.} - \Delta E$ (see fig.~\ref{fig:qmd}(b)),
the fusion barrier becomes
\begin{align}
	V_B = \frac{\Delta E}{1-\alpha(\kappa)}.\label{equ:fusion_barrier}
\end{align}
For \buck{}+\buck{} collisions, this approximate expression gives $V_B= 82$~eV,
which is very close to the exact model value ($V_B= 85$~eV) obtained in the upper dynamical calculations.
To obtain a first insight about the qualitative trends for the other combinations
(\buck{}+\rug{}, \rug{}+\rug{}), we ignore subtleties and use the same $k$ and $\Delta E$ values as obtained already for \buck{}+\buck{} by NA-QMD fine tuning.
The internal spring constant for \rug{}, however, is carefully chosen and fixed again by experiment.
From the (partly) non-degenerate five $H_g(1)$ modes in \rug{} ($E_2'$, $E_1''$, $A_1'$)  an experimental
mean value of $\omega_\text{mode}(\text{C}_{70}) = 261\ \text{cm}^{-1}$ has been
reported~\cite{Wang95} giving the spring constant $k_\text{mode}(\text{C}_{70}) = 0.55$~a.u.
For the (slightly) asymmetric \buck{}+\rug{} collision a reasonable mean value of $k_\text{mode}(\text{C}_{60})$ and $k_\text{mode}(\text{C}_{70})$ of $k_\text{mode}(\text{C}_{60}/\text{C}_{70}) = 0.53$~a.u. is used.

With these parameters, the predicted fusion barriers from (\ref{equ:fusion_barrier}) are compared with high precision QMD values~\cite{Knospe96}, in table~\ref{tab:fusion_barriers} (first two rows).
The analytical model reproduces the right trend and delivers absolute values within 10\% accuracy.
Obviously, the $H_g(1)$ frequencies $\omega_\text{mode}$ and fullerene masses $M$, eq.~(\ref{equ:kappa}) determine the fusion barriers in fullerene-fullerene collisions.

This is strongly supported by comparing the predictions with experimental data.
In this case, the finite temperature ($T \approx 2000$~K) of the colliding fullerenes has to be taken into account~\cite{Rohmund96_jpb}.
This has been done in the former QMD calculations~\cite{Knospe96} and led to the (well known) perfect agreement with the experimental data (cf. third and fifth rows in table~\ref{tab:fusion_barriers}).
In the present model, a finite temperature can be naturally taken into account by reducing the mode frequencies $\omega_\text{mode}$.
Using an arbitrary common scaling factor for $k_\text{mode}$ of $0.85$, the predicted fusion barriers by eq.~(\ref{equ:fusion_barrier}) are in beautiful agreement with the experiment (fourth and fifth row in table~\ref{tab:fusion_barriers}).
\begin{table}[t]
    \caption{Fusion barriers $V_B$ in~eV for various fullerene-fullerene combinations as predicted by our former QMD calculations~\cite{Knospe96} (first and third rows) and the present analytical model (second and fourth rows).
    The experimental values~\cite{Rohmund96_jpb} are presented in the last row.}

    \begin{tabular}{c|cc|cc|c}
   				& \multicolumn{2}{|c|}{$T=0$~K}			& \multicolumn{3}{|c}{$T=2000$~K}  	\\
				& QMD	& model 	& QMD   & model	& exp. \\ 
				 \hline
        \buck+\buck             & 80				& 82		& 60		& 65			& 60$\pm$1 \\ 
	\buck+\rug              & 94				& 87		& 70		& 69			& 70$\pm$6.5 \\ 
        \rug+\rug               & 104       			& 93		& 75		& 73			& 76$\pm$4 
    \end{tabular}
    \label{tab:fusion_barriers}
\end{table}

\section{Conclusion}
To summarize, \emph{ab initio} NA-QMD studies have finally cleared up the fusion mechanism in fullerene-fullerene collisions.
The  excitation of a ``giant'' $H_g(1)$ mode explains both large fusion barriers and small fusion cross sections.
This \emph{microscopic} picture is non-ambiguously confirmed by a \emph{macroscopic} spring model which depicts clearly the physics, reproduces the NA-QMD results and the experimental fusion barriers \emph{quantitatively}.

We note finally, that a ``giant'' vibrational excitation of the $A_g(1)$ breathing mode in \buck{} has been found recently in a time-resolved laser experiment~\cite{Laarmann07}.
The general investigation of large amplitude motion in fullerenes, including laser-induced \emph{fission}~\cite{Fischer2013}, could become an interesting new field of research.

\section{Appendix: Analytical solution in the entrance channel}
\renewcommand{\theequation}{A.\arabic{equation}}
\setcounter{equation}{0}

Inserting the harmonic potentials $U(R_c)$ and $V(D)$, the EOM (\ref{equ:newton}) can be written as
\begin{equation}\label{equ:eom}
  \begin{split}
        \mu \ddot R &= \left . -k\ (R_c-R_c^{0})\right |_{R_c=R-D} \\
	\mu_\text{mode} \ddot D &= -k_\text{mode}\ (D-D^0) + \frac{k}{2}\ \left . \left( R_c-R_c^{0} \right )\right |_{R_c=R-D}
  \end{split}
\end{equation}
which can be solved by making an exponential ansatz $e^{\imath \Omega t}$.
Doing so, the fundamental eigenfrequencies $\Omega_{1/2}$ of the system read
\begin{align}
    \Omega_{1/2} &= \omega\ \underbrace{\sqrt{\kappa + 1 \pm \sqrt{\kappa^2+1}}}_{f_{1/2}(\kappa)} \label{equ:freq1}
\end{align}
with the frequency $\omega = \sqrt{\frac{k}{\mu}}$ and the force constant ratio $\kappa$ as defined in eq.~(\ref{equ:kappa}).\\
With the initial conditions
\begin{align*}
R(0) &\equiv R^{0} = D^{0}+R_{c}^0,& \dot R(0) &= v_\text{c.m.},& \\
D(0) &= D^{0},& \dot D(0) &= 0& 
\end{align*}
the solution of (\ref{equ:eom}) is given by
\begin{align}\label{equ:analyt_solution1}
	R(t) &= \frac{v_\text{c.m.}}{\omega} \sum_{i=1,2} a_i \sin{(\Omega_i t)} +  R^{0},\\
	D(t) &= \frac{v_\text{c.m.}}{\omega} \sum_{i=1,2} b_i \sin{(\Omega_i t)} + D^{0}\label{equ:analyt_solution2}
\end{align}
with the amplitudes $a_{1/2} = \frac{1}{2\ f_{1/2}(\kappa)}\ (1 \mp \frac{\kappa}{\sqrt{\smash[b]{\kappa^2+1}}})$
and $b_{1/2} = \mp \frac{1}{2\ f_{1/2}(\kappa)}\ \frac{1}{\sqrt{\smash[b]{\kappa^2+1}}}$.

With the analytical solution (\ref{equ:analyt_solution1}), (\ref{equ:analyt_solution2}) 
the potential energy at the returning point $E^\text{ret}$ can be calculated.
$E^\text{ret}$ is given by
\begin{align}
E^\text{ret} &\equiv U(R_c^\text{ret}) = \frac{k}{2} \left(R_c^\text{ret}-R_{c}^0\right)^2\label{equ:e_ret}
\end{align}
with $R_c^\text{ret} \equiv R_c(t^\text{ret})=R(t^\text{ret})-D(t^\text{ret})$.\\
The returning time $t^\text{ret}$ defined by~$\dot R(t^\text{ret})=0$ is approximated by $t^\text{ret} \approx \frac{\pi}{2 \Omega_2}$. %
The approximation is justified since the second term of the sum in eq.~(\ref{equ:analyt_solution1}) dominates for the parameter range of $\kappa$ used here ($\frac{a_1}{\smash[t]{a_2}} \ll 1$).\\
Inserting the analytical solution (\ref{equ:analyt_solution1}), (\ref{equ:analyt_solution2}) in the definition (\ref{equ:e_ret}) and using the relation (\ref{equ:freq1})
we find 
\begin{align*}
	E^\text{ret} &= \alpha(\kappa)\ E_\text{c.m.}
\end{align*}
with the impact energy $E_\text{c.m.}=\frac{\mu}{2}\ v_\text{c.m.}^2$ and the coefficient $\alpha(\kappa)$ 
as defined before in eq.~(\ref{equ:alpha}).

\acknowledgments

We thank
Sebastian Schmidt (ETH Zurich) for many useful
comments and critical reading of the manuscript.
We gratefully acknowledge the allocation of computer resources from ZIH and MPI-PKS, Dresden and appreciate
the support of the DFG through Einzelverfahren.

\end{document}